\documentclass[journal, twocolumn]{IEEEtran}

\usepackage{algorithm}
\usepackage{algorithmic}
\usepackage{amsmath,amssymb,mathrsfs}
\usepackage{array}
\usepackage{balance}
\usepackage{booktabs}
\usepackage{cite}
\usepackage{fancyhdr}
\usepackage{float}
\usepackage[nomain,acronym,automake,toc,nopostdot]{glossaries}%[nonumberlist,acronym,toc]
\usepackage{graphicx,color}
\usepackage{mathtools}
\usepackage{multirow}
\usepackage{psfrag}
\usepackage{stfloats}
\usepackage[justification=centering]{caption}

\usepackage{subfigure}
\usepackage{float}
\makeglossaries
\newacronym{awgn}{AWGN}{additive white Gaussian noise}
\newacronym{hts}{HTS}{high throughput satellites}
\newacronym{3gpp}{3GPP}{3rd generation partnership project}
\newacronym{mmse}{MMSE}{minimum mean square error}
\newacronym{ls}{LS}{least square}
\newacronym{ofdm}{OFDM}{orthogonal frequency division multiplexing}
\newacronym{lstm}{LSTM}{long short-term memory}
\newacronym{cnn}{CNN}{convolutional neural network}
\newacronym{csi}{CSI}{channel state information}
\newacronym{cs}{CS}{compressed sensing}
\newacronym{dnn}{DNN}{deep neural netrowk}
\newacronym{elm}{ELM}{extreme learning machine}
\newacronym{slfn}{SLFN}{signal hidden layer feed-forward network}
\newacronym{celm}{CELM}{complex-valued ELM}
\newacronym{hpa}{HPA}{high power amplifer}
\newacronym{qam}{QAM}{quadrature amplitude modulation}
\newacronym{wlp}{WLP}{widely linear processing}
\newacronym{racelm}{RACELM}{regularized augmented CELM}
\newacronym{celmai}{CELMAI}{CELM with augmented input}
\newacronym{celmah}{CELMAH}{CELM with augmented hidden layer}
\newacronym{twta}{TWTA}{travelling wave tube amplifiers}
\newacronym{sspa}{SSPA}{solid-state power amplifiers}
\newacronym{obo}{OBO}{output back off}
\newacronym{imux}{IMUX}{input multiplexer}
\newacronym{omux}{OMUX}{output multiplexing filter}
\newacronym{am}{AM}{amplitude modulation }
\newacronym{pm}{PM}{phase modulation}
\newacronym{papr}{PAPR}{peak-to-average power ratio}
\newacronym{leo}{LEO}{low earth orbit}
\newacronym{nlos}{NLOS}{non-line-of-sight}
\newacronym{tdl}{TDL}{tapped delay line}
\newacronym{rf}{RF}{radio frequency}
\newacronym{r-celmah}{R-CELMAH}{regularized CELMAH}
\newacronym{lms}{LMS}{least mean square}
\newacronym{mse}{MSE}{mean square error}
\newacronym{ber}{BER}{bit error rate}
\newacronym{snr}{SNR}{signal-to-noise ratio}
\newacronym{ibo}{IBO}{input back-off}
\newacronym{dft-s-odfm}{DFT-s-OFDM}{DFT-spread-OFDM}
\newacronym{iq imbalance}{I/Q imbalance}{inphase and quadrature signals imbalance}
\newacronym{nr}{NR}{new radio}
\newacronym{b5g}{B5G}{beyond 5G}
\newacronym{ppm}{ppm}{parts per million}
\newacronym{celmwlls}{CELM-WLLS}{CELM augmented by WL-LS}
\newacronym{isi}{ISI}{inter-symbol interference}
\newacronym{wlls}{WL-LS}{widely linear least square}
\newacronym{fft}{FFT}{fast fourier transform}
\newacronym{pdf}{PDF}{probability density function}

\definecolor{sblue}{RGB}{0,51,120}
\definecolor{sred}{RGB}{200,51,130}

\renewcommand{\eqref}[1]{(\ref{#1})}

\newcommand{\mysize}{8.5cm}

\ifCLASSINFOpdf
\else
\fi

\begin{document}
\title{Widely Linear Augmented Extreme Learning Machine Based Impairments Compensation for Satellite Communications}
\author{Yang Luo,
	Arunprakash Jayaprakash,
	Gaojie Chen, \textit{Senior Member, IEEE},
	Chong Huang, \textit{Member, IEEE}, \\
	Qu Luo, \textit{Member, IEEE}, Pei Xiao, \textit{Senior Member, IEEE}
{\thanks{\small Y. Luo, C. Huang, Q. Luo and P. Xiao are with 5GIC and 6GIC, Institute for Communication Systems, University of Surrey, Guildford, UK, GU2 7XH. A. Jayaprakash is with Satellite Applications Catapult, Harwell Oxford, UK, OX11 0QR. G. Chen is with School of Flexible Electronics (SoFE) \& State Key Laboratory of Optoelectronic Materials and Technologies, Sun Yat-sen University, Guangdong, 510220, China. Email: (yl01961@surrey.ac.uk, arunprakash.jayaprakash@sa.catapult.org.uk, gaojie.chen@ieee.org, chong.huang@surrey.ac.uk, q.u.luo@surrey.ac.uk, p.xiao@surrey.ac.uk)}}
}
\markboth{}%
{}

\maketitle

\begin{abstract}
Satellite communications are crucial for the evolution beyond fifth-generation networks. However, the dynamic nature of satellite channels and their inherent impairments present significant challenges. In this paper, a novel post-compensation scheme that combines the complex-valued extreme learning machine with augmented hidden layer (CELMAH) architecture and widely linear processing (WLP) is developed to address these issues by exploiting signal impropriety in satellite communications. Although CELMAH shares structural similarities with WLP, it employs a different core algorithm and does not fully exploit the signal impropriety. By incorporating WLP principles, we derive a tailored formulation suited to the network structure and propose the CELM augmented by widely linear least squares (CELM-WLLS) for post-distortion. The proposed approach offers enhanced communication robustness and is highly effective for satellite communication scenarios characterized by dynamic channel conditions and non-linear impairments. CELM-WLLS is designed to improve signal recovery performance and outperform traditional methods such as least square (LS) and minimum mean square error (MMSE). Compared to CELMAH, CELM-WLLS demonstrates approximately 0.8~dB gain in BER performance, and also achieves a two-thirds reduction in computational complexity, making it a more efficient solution.
\end{abstract}
\glsresetall

\setlength{\parindent}{0pt}

\section{Introduction}\label{sec01}
Satellite communications have attracted much attention due to the emergence of \gls{hts} and non-geo-stationary-orbit satellite constellations, which are pivotal components for the evolution beyond fifth-generation networks. The integration of \gls{b5g} with satellites has been highlighted for delivering broadband services in areas with limited coverage and addressing specific needs such as marine, aerial, and rail communication \cite{jiang2023software}. Notably, the deployment details of non-terrestrial networks, including associated satellite communication channel models at the physical layer, key parameters, and necessary adjustments for \gls{nr} standards, are outlined in \cite{10179219}. This work leans on the frameworks provided by the above literature, indicating that the successful integration of satellite systems requires overcoming challenges such as reducing latency, enhancing spectral efficiency, and mitigating satellite channel impairments \cite{guidotti2019architectures}.

Link budget constraints are critical in the physical layer design for the downlink of satellite communications and are exacerbated by the inherent impairments of satellite channels \cite{huang2024joint}. Particularly, specialized mitigation measures are needed, such as pre-distortion at the transmitter or post-distortion at the receiver. However, pre-distortion often requires prior knowledge of the non-linear characteristics or additional feedback loops, which complicates satellite system design  \cite{9792424}. Considering the link budget constraints for satellite communications,  pre-distortion approaches are less feasible, and post-distortion has become a main research focus. In \cite{10246293}, a pilot-based post-distortion approach was devised for the channel estimation of \gls{leo}. Furthermore, \cite{10066300} studied a pilot-assisted Bayesian posterior channel estimator based on the orthogonality of the pilot. While these techniques provide post-distortion solutions suited to satellite hardware limits and demonstrate the feasibility of a pilot-based scheme, they still need to grapple with the drawback of high computational intensity. In contrast, machine learning (ML) methods are emerging as effective tools against satellite impairments distortion-induced signal interference, with the added advantage of lower online computational complexity \cite{9944375}.

ML has shown tremendous potential in channel estimation \cite{10273640}. However, conventional ML methods face several challenges when applied to satellite communication scenarios. Specifically, the performance of traditional ML approaches is heavily influenced by the accuracy of the training labels, which means the accuracy of label collection directly impacts performance \cite{9999531}. Considering the dynamic nature of satellite environments is characterized by frequent changes and multiple impairments, it necessitates a large amount of computationally complex data collection \cite{10328395}. Additionally, the intricate satellite channels lead to noisy labels or complex label distribution, which could cause the gradient descent method to become unstable due to its susceptibility to randomness, thereby reducing overall performance \cite{algan2021image}. Furthermore, the conventional one-go model of traditional ML typically would not involve adjusting weights and biases after the pre-training, and the dynamics of actual channel environments frequently deviate from these predefined models. Specifically, the learning algorithms might be optimized and fitted with the pre-training set rather than actual channel conditions, resulting in performance degradation if encountering out-of-distribution data, particularly in dynamic environments \cite{9546673}. Therefore, exploring the network structures that are more suitable for satellite scenarios is necessary.

Given the inherent impairments in satellite systems, \gls{elm} has demonstrated strong performance in handling hardware impairments \cite{9310357}. As a type of \gls{slfn}, ELM has been theoretically proven to possess the universal approximation property, which ensures its capacity to approximate any measurable function given sufficient hidden nodes \cite{6125249}. This foundational property supports the use of ELM-based architectures in signal estimation tasks under complex channel conditions. Furthermore, with the requirements of maximizing satellite payload efficiency and minimizing additional latency, ELM, as proposed in \cite{csahin2014application}, has proven to be a promising lightweight, low-complexity solution for satellite communications. It strikes a good balance between accuracy and adaptability to varying channel conditions. Specifically, as a post-distortion online training network,  ELM directly uses the pilot and received signals as the training set, which reduces the accuracy requirement of the label collection for neural network training\cite{8715649}. In addition, ELM's parameter optimization schemes have a fast and straightforward structure. Compared to traditional parameter optimization schemes, such as gradient descent, ELM is less affected by the distribution of the training set and avoids convergence to the local optimization. However, the calculation of ELM weight parameters between the hidden and output layer is generally solved using least squares, targeting the pilot signal to find the current optimization parameters, without the need to achieve global optimal of communication \cite{soares2016adaptive}. Moreover, ELM's efficient parameter-solving mechanism and training set acquisition enable rapid adaptation to dynamic environments \cite{javed2015new}. To further enhance adaptability, Domain Adaptation ELM (DAELM) has been proposed to address distribution discrepancies between source and target domains using transfer learning, enabling rapid adaptation to new case based on existing data \cite{zhang2014domain,chen2018domain}. However, traditional ELM-based methods predominantly rely on least squares estimation for weight calculation, which is inherently limited in handling data impropriety. Given the impairments specific to satellite communication, particularly the rotational variant effect that induces signal impropriety, there is further potential for improving ELM-based approaches to better exploit the statistical properties of improper signals.

The rotational variant is a common phenomenon in the satellite communication scenario, and it gives rise to the exploitable additional degree of freedom due to the impropriety of the received signals \cite{9505614}. From a signal transmission point of view, the authors analyzed the rotational invariance in \cite{5961645} and found the probability distribution of the \gls{qam} symbols is no longer valid in the presence of satellite channel distortions, which results in a non-circular received constellation. However, massive researches of ELM channel estimation approach are based on the assumption that the complex received signal has equal energy and uncorrelated real-imaginary parts. While existing research recognizes and emphasizes the potential of signal impropriety for enhanced signal processing, its full utilization remains an open area for further investigation \cite{chen2021widely}. To elaborate further, the impropriety of the signal is caused by the correlation between a complex random variable and its conjugate, called pseudo-autocorrelation, which can be dealt with as an additional statistical characterization \cite{5961645}. In particular, the ability and effectiveness of \gls{wlp} to exploit additional statistical characteristics showed robustness under different \gls{csi} conditions in channel estimation, which demonstrated WLP as an optimal strategy for managing rotational variant complex random processes \cite{zhang2021adaptive}. To enhance the performance of ELM, the authors in \cite{zhang2018augmented} proposed \gls{celmah}, this model integrates the concept of widely linear concept with \gls{celm}. This structure is well-suited for satellite channels with signal impropriety, which aligns naturally with the strengths of widely linear processing. However, \gls{celmah} strictly adheres to the least square method for weight matrix calculation as traditional ELM, fundamentally differing from the WLP approach. Therefore, to tackle impropriety and impairments in satellite scenarios, applying widely linear processing should be appropriately considered with the specific formula derivation for better potential and performance.

To address the impairments and signal impropriety present in satellite communications, we introduce a novel approach based on augmented-CELM, which integrates the functions of channel estimation and equalization. An online learning strategy is adopted, utilizing pilot and received signals to provide the initial training set for ML as preliminary CSI knowledge. Furthermore, to evaluate the reliability of augmented-CELM in satellite channels, we derive the widely linear weight calculation for the first time. The main contributions of this paper are summarized as follows:
\begin{itemize}
	\item A post-distortion method for satellite down-link channels is proposed, which combines ELM and WLP for the first time. This approach enhances the weight calculation between the hidden and output layers using WLP, making it more suitable for the satellite impairment scenarios and network architectures.
	
	\item A theoretical analysis of WLP within the CELM framework is presented, where the expectation is replaced with instantaneous information in widely linear operations, allowing the use of pseudo-autocorrelation as an additional second-order statistical feature.
	
	\item The \gls{celmwlls} post-distortion approach proposed in this work offers a lightweight solution, achieving two-thirds the FLOPs computational complexity of the traditional \gls{celmah} method, while delivering improved performance.
	
	\item Multiple factors influencing satellite channels, including inherent impairments and rotational invariance that cause signal impropriety, are considered. Compared to previous studies, the satellite channel model in this work is more comprehensive and better reflects the practical channel conditions.
\end{itemize}
The remainder of the paper is organized as follows. A brief overview of satellite communication transceiver and channel payload impairments is presented in section \ref{sec02}. CELMAH and \gls{racelm} based receiver post-distorter are introduced in Section \ref{sec03}. The numerical analysis and simulation results to verify the effectiveness of post-distorter in the satellite scenario are presented in Section \ref{sec04} and concluding remarks are drawn in Section \ref{sec05}.
\begin{figure*}[htbp]
	\centering
	\includegraphics[scale=0.85]{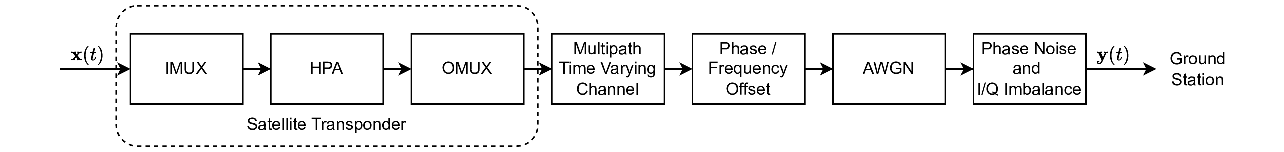}
	\caption{Satellite system impairments}
	\label{Satellite_Channel}
\end{figure*}

%%%%%%%%%%%%%%%%%%%%%%%%%%%%%%%%%%%%%%%%%%%%%%%%%%%%%%%%%%%%%%%%%%%

\section{Satellite System Impairments}\label{sec02}
This section provides a concise overview of the impairments affecting the end-to-end satellite transceiver and payload within the integrated satellite to ground station communication system architecture. It focuses mainly on the base-band model of the satellite channel, highlighting distortions caused by both the transceiver and the radio channel. Fig.~\ref{Satellite_Channel} illustrates the end-to-end block diagram of point-to-point satellite-ground integration. Detailed discussions of the impacts mentioned are presented in the sequel.

\subsection{High Power Amplifier}
A key impairment in satellite transmission efficiency is the high non-linearity of the onboard \gls{hpa}, where \gls{twta} is widely used. The HPA is driven close to saturation to maximize power efficiency, significantly increasing non-linear distortion. Even with advanced pre-distortion techniques to mitigate non-linearity, avoiding clipping distortion requires operating the HPA with substantial \gls{obo}, depending on the \gls{papr} of the transmitted waveform. Moreover, jointly amplified signals through a single HPA cause inter-modulation distortion in multi-carrier systems. The satellite transponder setup, depicted in Fig.~\ref{Satellite_Channel}, includes an \gls{imux} for signal selection, the non-linear HPA, and an \gls{omux} to minimize interference. The HPA’s amplitude and phase distortions are defined by its \gls{am} and \gls{pm} conversion responses. The input of HPA can be expressed by polar coordinates as $\mathbf{x}(t) = A(t)e^{j\theta(t)}$, where $A(t)$ and $\theta(t)$ are amplitude and phase of input signal respectively. Therefore, the base-band equivalent output can be $\mathbf{v}(t)$ represented as \cite{jayaprakash2020new}{\color{blue},}
\begin{equation}
	\mathbf{v}(t)=T(A(t)) e^{j((\theta(t))+\phi(A(t)))} ,
	\label{equ:HPA}
\end{equation}
\noindent where $T(A(t))$ and $\phi(A(t))$ are the AM/AM and AM/PM responses of the HPA respectively. For a typical HPA, the relationship between the input and output and the AM/AM and AM/PM characteristics can be modelled with generic memoryless modified Saleh model \cite{o2009new} or a polynomial model.
\subsection{Doppler Shift}
Doppler effects significantly impair non-geostationary satellite-based integrated satellite-terrestrial communications more than in terrestrial scenarios. These effects stem from the high relative velocity between LEO satellite transmitters and receivers. The \gls{nlos} propagation channel between satellite and terrestrial links resembles terrestrial multi-path propagation but with nearly zero angular spread from the satellite. The \gls{tdl} model, which outlines the power delay profiles for multi-path channels at different satellite elevation angles (for frequencies below 6 GHz) and the required Doppler shift for each tap, which can be used to simulate the channel propagation effects in the integrated satellite terrestrial scenario \cite{3gpp2019study}. Let $h(\tau,t)$ represent the time-varying impulse response of the channel, and give the channel-affected signal as
\begin{equation}
	\tilde{\mathbf{v}}(t)=\sum_{\mathrm{i}=1}^L h_i \mathbf{v}\left(t-\tau_i\right) e^{-j 2 \pi (f_0+f_d(t))(t- \tau_i)+\sigma_{0}} + n\left(t\right),
	\label{equ:timevarying impulse response}
\end{equation}
\noindent where $L$ is the number of the taps, $f_0$ is the carrier frequency, $\tau_i$ is the delay of the $\mathrm{i}$th tap, $\sigma_{0}$ is the phase offset of the up-converter oscillator, and $n\left(t\right)$ is \gls{awgn}. The Doppler distortions can be pre-compensated by the knowledge of the serving satellite ephemeris and velocity state vectors. A closed-form expression, provided in \cite{ali1998doppler}, to calculate the time-varying Doppler shift $f_d (t)$ for a satellite is given as
\begin{equation}
	\mathbf{f}_{d}(t)=-\frac{f_{c} w_{s} r_{e} r_{o} \sin \left(w_{s} t\right) \eta\left(\theta_{\max }\right)}{c \sqrt{r_{e}^{2}+r_{o}^{2}-2 r_{e} r_{o} \cos \left(w_{s} t\right) \eta\left(\theta_{\max }\right)}} ,
	\label{equ:dopplershift}
\end{equation}
\noindent where $\eta(\theta_{\max})=\cos \left(\cos^{-1}\left(\frac{r_{e}}{r_{o}} \theta_{\max}\right)-\theta_{\max}\right)$, and $w_{s}$, $\theta_{\max}$, $r_o$, $r_e$ represent the angular velocity, maximum elevation angle, satellite orbit radius, and Earth's radius, respectively.
\subsection{Frequency Offest}
Users near the beam edge still experience a residual frequency offset, which shifts the center frequency of the base-band signal, leading to severe frequency synchronization issues \cite{jayaprakash2020new}. Representing the residual frequency offset as $\epsilon$, the frequency-shifted signal can be expressed by
\begin{equation}
	\mathbf{\tilde{v}}_{d}(t)=\mathbf{\tilde{v}}(t) e^{-j 2 \pi \epsilon t} .
	\label{equ:dopplershift1}
\end{equation}
\subsection{Phase Noise and I/Q Imbalance}
Phase noise and \gls{iq imbalance} are significant \gls{rf} hardware impairments in satellite communication links. Phase noise, resulting from variations in the oscillators of the transmitter and receiver, spreads power across adjacent frequencies, causing spectrum regrowth and random rotation of received constellations. In contrast, I/Q imbalance, due to mismatches in the oscillators that generate in-phase and quadrature components during signal conversion, similarly disrupts the integrity of the received signal. In \gls{ofdm} systems, phase noise disrupts the orthogonality of subcarriers, leading to inter-carrier interference.  Let $\tilde{v}_{d}(t)=\tilde{v}_{d}^{I}(t)+j \tilde{v}_{d}^{Q}(t)$. The complex-valued base-band received signal $\mathbf{y}\left(t\right)$, affected by I/Q imbalance \cite{sharma2020system} and phase noise \cite{khanzadi2014calculation} can be expressed as
\begin{equation}
	\begin{aligned}
		\mathbf{y}(t)= & {\left[\left(1+\eta_{A}\right)\left(\mathbf{x}_{I}(t) \cos \left(\frac{\eta_{\phi}}{2}\right)-\mathbf{x}_{Q}(t) \sin \left(\frac{\eta_{\phi}}{2}\right)\right)+\right.} \\
		& \left.j\left(1-\eta_{A}\right)\left(\mathbf{x}_{Q}(t) \cos \left(\frac{\eta_{\phi}}{2}\right)-\mathbf{x}_{I}(t) \sin \left(\frac{\eta_{\phi}}{2}\right)\right)\right] e^{j \phi(t)} ,
	\end{aligned}
\end{equation}
\noindent where $j$ is the imaginary parameter of a complex number. $\frac{\eta_A}{2}$ and $\frac{\eta_\phi}{2}$ denote the amplitude and phase imbalances respectively. Additionally, phase noise originates from phase fluctuations in the local oscillators of both the transmitter and receiver, introducing undesired distortions in the received signal. These fluctuations can be modeled as a stochastic process $\phi(t)$, resulting in a multiplicative distortion $e^{j \phi(t)}$, which effectively characterizes the phase noise according to the S-band satellite phase-noise mask \cite{khanzadi2014calculation}. In practical satellite communication systems, phase noise spreads power across adjacent frequencies, causing spectrum regrowth, disrupting carrier synchronization, and degrading signal demodulation accuracy. Similarly, I/Q imbalance, arising from mismatches in the in-phase and quadrature signal paths, introduces additional distortion in the received signal. Together, these impairments alter the statistical properties of the transmitted signal, transforming originally circularly symmetric symbol constellations into non-circular received signals, a distortion that necessitates robust compensation techniques at the receiver \cite{javed2020journey}.
\begin{figure}[t!]
	\centering
	\includegraphics[scale=0.9]{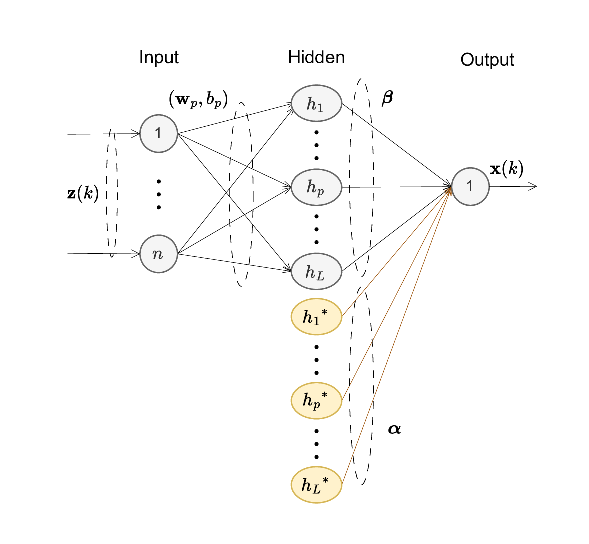}
	\caption{Architecture of the augmented CELM}
	\label{fig_celmah}
\end{figure}

%%%%%%%%%%%%%%%%%%%%%%%%%%%%%%%%%%%%%%%%%%%%%%%%%%%%%%%%%%%%%%%%%%%
\section{Augmented Extreme Learning Machine Based Receiver Post-Distorter}\label{sec03}
To enhance the performance of CELM in satellite communication scenarios and utilize signal impropriety in this context, we propose a widely linear augmented complex-valued extreme learning machine-based receiver post-distorter for satellite communication systems. In this section, we first introduce the CELMAH models proposed in \cite{zhang2018augmented}, and the architecture of the network is shown in Fig.~\ref{fig_celmah}. Additionally, we will investigate the exploitation of signal impropriety resulting from non-linear satellite channel impairments and derive the relevant widely linear algorithms.
\subsection{CELM with augmented hidden layer}
CELMAH is based on CELM with augmented hidden layer, and the pilot signals are used as prior information for the pre-training in the satellite communications scenario. The transmitted signal
$\textbf{x}=\left[x_1, x_2, \ldots, x_N\right]^{T}\in\mathbb{C}^{\mathrm{N}}$. Considering $\textbf{y}$ is the aggregated received signal over $\mathrm{I}$-taps multi-path delay spread, then the received signal can be expressed as $\textbf{y}=\left[y_1, y_2, \ldots, y_{N+I-1}\right]^{T}\in\mathbb{C}$. Therefore, based on \gls{tdl} model, the $N$ samples for each $\mathrm{i}$th tap of all $\mathrm{I}$-taps can be derived as
\begin{equation}
	\begin{aligned}
		\textbf{z}_i & = \left[z_{i, 1}, z_{i, 2}, \ldots, z_{i, N}\right]^{T} \\
		& = \left[y_{1+I-i}, y_{2+I-i}, \ldots, y_{N+I-i}\right]^{T}
	\end{aligned}
\end{equation}

Given $N$ arbitrary distinct data samples $\{(\mathbf{z}(k), \mathbf{x}(k))\}_{k=1}^{N}$, where $\mathbf{x}(k)\in\mathbb{C}^{\mathrm{N}}$ present the target. $\mathbf{z}(k)\in\mathbb{C}^{\mathrm{N}\times\mathrm{I}}$ presents the input of the network, which stands for $\mathrm{N}$ samples with $\mathrm{I}$ features (multi-path taps). The column vector $\mathbf{w}_{p} \in \mathbb{C}^{N}$ contains the hidden layer input weights connecting the input layer neurons to the $p$th hidden neuron, and $b_{p}\in \mathbb{C}$ is the bias of the $p$th of all $L$ hidden nodes. These randomly initialized weights and biases operate the input data into the hidden layer nodes with a non-linear mapping function $g(\cdot)$. Therefore, the output of the hidden layer corresponding to the $p$th neuron is then given as
\begin{equation}
	\begin{aligned}
		h_p= g\left(\mathbf{w}_{p} \cdot \mathbf{z}\left(k\right)+b_{p}\right), \quad k=1,2, \ldots, N .
		\label{equ:hidden_node}
	\end{aligned}
\end{equation}
Where $\left( \cdot\right) $ represents the inner product. The network structure of the CELMAH has augmented hidden layer by an additional conjugate of the hidden nodes. Fig.~\ref{fig_celmah} describes the structure of the additional conjugate hidden nodes. Hence, the output matrix of the hidden layer can be represented as
\begin{scriptsize}
	\begin{equation}
		\setlength\arraycolsep{0.3pt} % 设置数组中列的间距
		\begin{array}{l}
			\Tilde{\mathbf{H}} = \left[\mathbf{H}, \mathbf{H}^{*}\right] \\
			= \left[\begin{array}{cccccccccc}
				h_{1}(\mathbf{z}(1)) & \cdots & h_{p}(\mathbf{z}(1)) & \cdots & h_{L}(\mathbf{z}(1)) & h_{1}^{*}(\mathbf{z}(1)) & \cdots & h_{p}^{*}(\mathbf{z}(1)) & \cdots & h_{L}^{*}(\mathbf{z}(1)) \\
				\vdots & & \ddots & & \vdots & \vdots & & \ddots & & \vdots \\
				h_{1}(\mathbf{z}(N)) & \cdots & h_{p}(\mathbf{z}(N)) & \cdots & h_{L}(\mathbf{z}(N)) & h_{1}^{*}(\mathbf{z}(N)) & \cdots & h_{p}^{*}(\mathbf{z}(N)) & \cdots & h_{L}^{*}(\mathbf{z}(N))
			\end{array}\right]_{N \times 2L .}
		\end{array}
		\label{equ:augmentedHiddennode}
	\end{equation}
\end{scriptsize}
The conjugate counterparts of the hidden nodes, as delineated in (\ref{equ:augmentedHiddennode}), can be defined as
\begin{equation}
		h^{*}_{p}= {g\left(\mathbf{w}_{p} \cdot \mathbf{z}\left(k\right)+b_{p}\right)}^{*}={x\left(k\right)}^{*}, \quad k=1,2, \dots, N.
\end{equation}
The augmented weights matrix, which links the hidden layer and the output layer is denoted as
\begin{equation}
	\boldsymbol{\Tilde{\beta}}=\left[\boldsymbol{\beta}^{T}, \boldsymbol{\alpha}^{T}\right]^{T}.
	\label{augmented_hiddenlayer}
\end{equation}
\noindent where $\boldsymbol{\beta}$ establishes a connection from the hidden nodes $h_1$ through $h_L$ to the output layer, while $\boldsymbol{\alpha}$ links the conjugated hidden nodes $h_{1}^{*}$ through $h_{L}^{*}$ to the output layer
\begin{equation}
	\boldsymbol{\beta}=\left[\begin{array}{c}
		\beta_{1}^{T} \\
		\vdots \\
		\beta_{L}^{T}
	\end{array}\right]_{L \times 1} \text{and } \boldsymbol{{\alpha}}=\left[\begin{array}{c}
		{\alpha}_{1}^{T} \\
		\vdots \\
		{\alpha}_{L}^{T}
	\end{array}\right]_{L \times 1} .
\end{equation}
\noindent Therefore, the estimated output $\hat{\textbf{x}}$ is represented as
\begin{equation}
	\hat{\mathbf{x}}= \mathbf{H}\boldsymbol{\beta}^{H}+ \mathbf{H}^{*}\boldsymbol{\alpha}^{H} .
	\label{CELMAH_h2o}
\end{equation}
When the estimate is towards to the target vector $\mathbf{x}$, (\ref{CELMAH_h2o}) can be simplified to
\begin{equation}
	\Tilde{\mathbf{H}} \Tilde{\boldsymbol{\beta}}=\mathbf{x} .
	\label{hiddentooutput}
\end{equation}
In general, the accurate solution $\Tilde{\boldsymbol{\beta}}$ of (\ref{hiddentooutput}) does not exist. Instead, it is solved with the least-squares solution given by
\begin{equation}
	{\Tilde{\boldsymbol{\beta}}}=\left(\Tilde{\mathbf{H}}\right)^{\dagger} \mathbf{X},
	\label{lesat_square}
\end{equation}
\noindent where $(\Tilde{\mathbf{H}})^{\dagger}$ represents the Moore-Penrose generalized inverse of the matrix $\Tilde{\mathbf{H}}$. The training process of the CELMAH is summarized by \textbf{Algorithm \ref{algorithm:celmah}}.
\begin{algorithm}[!t]
	\caption{CELMAH}
	\begin{algorithmic}[1]
		\STATE \textbf{Procedure} CELMAH\_Train($\mathbf{Z}, \mathbf{x}$)
		\STATE Randomly initialize input weights $\mathbf{W}$ and biases $\mathbf{b}$
		\STATE Compute hidden layer output $\mathbf{H} = g(\mathbf{W} \cdot \mathbf{Z} + \mathbf{b})$
		\STATE Augment the hidden layer by complex conjugate $\mathbf{\Tilde{H}} = [\mathbf{H}, \mathbf{H}^{*}]$
		\STATE Compute output weights  ${\Tilde{\boldsymbol{\beta}}}=\left(\Tilde{\mathbf{H}}\right)^{\dagger} \mathbf{X}$ for $\mathbf{\Tilde{H}}$
		\STATE \textbf{Return} $\mathbf{W}$, $\mathbf{b}$, and ${\Tilde{\boldsymbol{\beta}}}$
	\end{algorithmic}
	\label{algorithm:celmah}
\end{algorithm}
\begin{figure}[!t]
	\includegraphics[width=4.5cm]{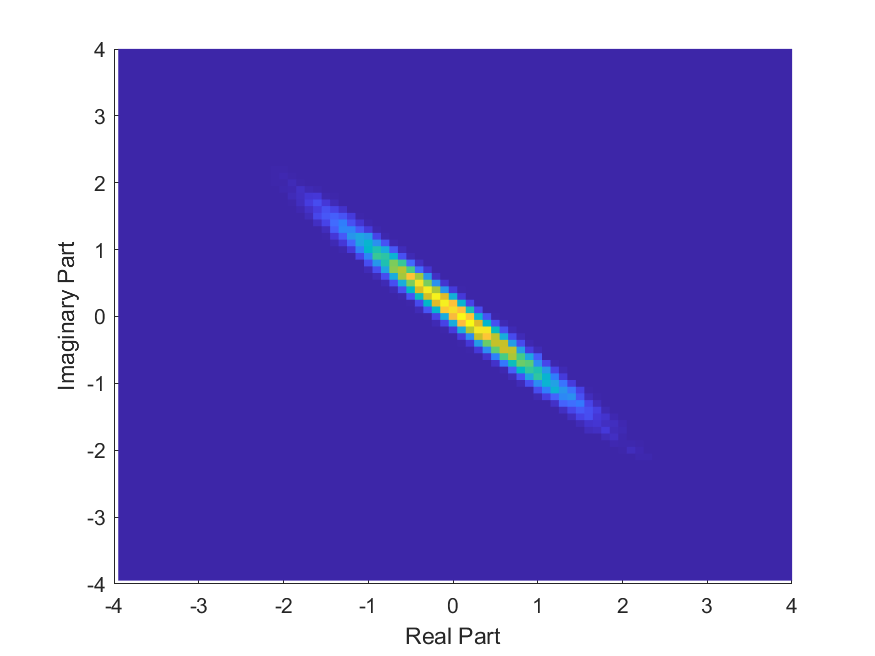}\includegraphics[width=4.5cm]{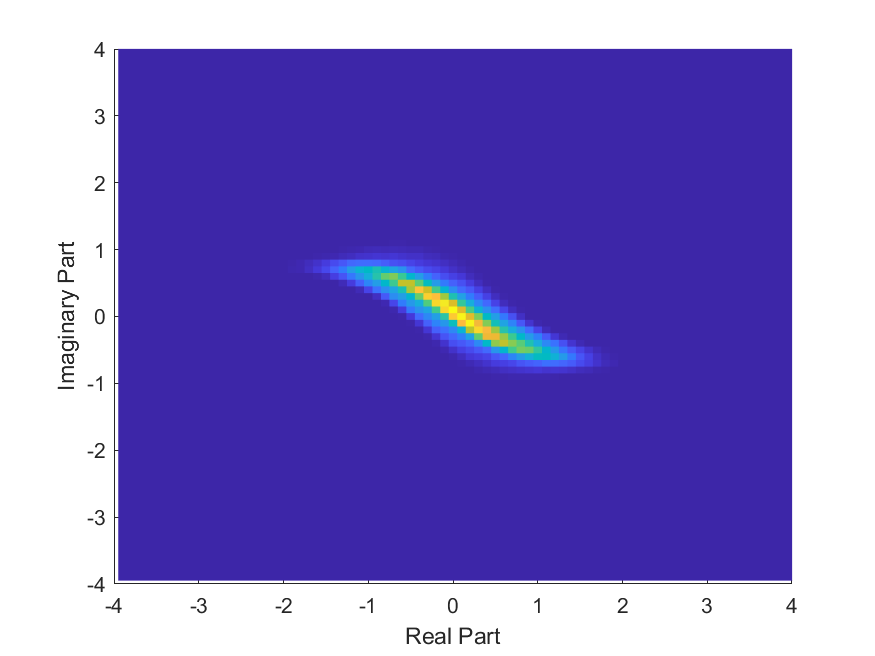}
	\caption{PDF comparison of received signal (left) and hidden layer output (right)}
	\label{fig_pdf}
\end{figure}

The left plot in Fig.~\ref{fig_pdf} shows a narrow \gls{pdf} for the received signal, highlighting the impact of impropriety. In contrast, the hidden layer output (right) displays a broader distribution due to ML generalization and deforms due to the nonlinear activation. Despite this, the hidden layer output retains non-zero pseudo-autocorrelation, a key condition for applying WLP \cite{9635690}. However, simply applying \gls{ls} to the augmented matrix $\Tilde{\mathbf{H}}$ to obtain the weights $\boldsymbol{\beta}$ and $\boldsymbol{\alpha}$ does not necessarily allow for the proper exploitation of the additional second-order statistics, even though the network structure of the CELMAH model generates a linear output and is similar to the widely linear \gls{lms} algorithm. It is still necessary to analyze the weights calculation from the hidden layer to the output layer in CELMAH through the lens of WLP, and revisit this regression problem. Let $\mathbf{H}$ be a matrix random variable to be estimated based on an observation that is a random vector $\mathbf{x}$. The estimate $\mathbf{H}$ that minimizes the \gls{mse} is the regression or the conditional expectation value $\mathbb{E}[\mathbf{x}|\mathbf{H}]$. This result generally holds when $\mathbf{H}$ and $\mathbf{x}$ are real-valued. However, it remains valid for complex-valued quantities. For real $\mathbf{H}$ and $\mathbf{x}$, the regression is linear, and it is also linear for complex data in both $\mathbf{H}$ and $\mathbf{H}^*$, termed widely linear \cite{picinbono1995widely}. This analysis reveals that the mathematical model of WLP for weights calculation can be similarly formulated as (\ref{CELMAH_h2o}). Additionally, the CELMAH structure, akin to WLP, enables the exploration of the second-order statistics of signals, which can be assessed through the covariance matrix of the hidden layer's output vector. For the complex-valued vector $\mathbf{H}$, the autocorrelation and pseudo-autocorrelation are denoted as
\begin{equation}
	\mathcal{C}_{\mathbf{H H}}=\mathbb{E}\left[\mathbf{H}^{H} \mathbf{H}\right] \text { and } \mathcal{P}_{\mathbf{H H}}=\mathbb{E}\left[\mathbf{H}^{T} \mathbf{H}\right].
\end{equation}
\noindent Considering that the augmented weights matrix is defined in (\ref{equ:augmentedHiddennode}) as $\Tilde{\mathbf{H}} = \left[\mathbf{H},\mathbf{H}^{*}\right]$, the corresponding covariance matrix is as follows:
\begin{equation}
	\mathcal{C}_{\Tilde{\mathbf{H}} \Tilde{\mathbf{H}}} = \mathbb{E}\left[\Tilde{\mathbf{H}}^{H} \Tilde{\mathbf{H}}\right] = \mathbb{E}\Bigg[\left[\begin{smallmatrix}
		\mathbf{H}^{H} \\ \mathbf{H}^{T}
	\end{smallmatrix}\right] \left[\mathbf{H}, \mathbf{H}^{*}\right] \Bigg]= \left[\begin{smallmatrix}
		\mathcal{C}_{\mathbf{H H}} & {\mathcal{P}_{\mathbf{H H}}}^{*} \\
		\mathcal{P}_{\mathbf{H H}} & {\mathcal{C}_{\mathbf{H H}}}^{*}
	\end{smallmatrix}\right].
\label{equ:celmah_secondorder_statistics}
\end{equation}
\noindent However, the structure of CELMAH can not provide any expectation for autocorrelation and pseudo-autocorrelation. Therefore, it can not simply combine CELMAH with WLP directly, which is necessary to derive a suitable widely linear regressor based on the structure of CELMAH

\subsection{Widely linear augmented CELM}
\begin{algorithm}[t]
	\caption{CELM-WLLS}
	\begin{algorithmic}[1]
		\STATE \textbf{Procedure} CELM-WLLS\_Train($\mathbf{Z}, \mathbf{x}$)
		\STATE Randomly initialize input weights $\mathbf{W}$ and bias $\mathbf{b}$
		\STATE Compute hidden layer output $\mathbf{H} = g(\mathbf{W} \cdot \mathbf{Z} + \mathbf{b})$
		\STATE Augment the hidden layer by complex conjugate $\mathbf{\Tilde{H}} = [\mathbf{H}, \mathbf{H}^{*}]$
		\STATE Calculate the widely linear augmented output weights matrix separately: \\ $\boldsymbol{\alpha} = \left( \mathbf{S}^{*}+\mathbf{R}\mathbf{C}^{-1}\mathbf{P}^{*} \right) \left( \mathbf{P}\mathbf{C}^{-1}\mathbf{P}^{*} + \mathbf{C}^{*} \right)^{-1}$ for $\mathbf{H}$\\ $\boldsymbol{\beta} = \left( \mathbf{R}+\mathbf{S}\mathbf{C}^{*-1}\mathbf{P} \right) \left( \mathbf{P}^{*}-\mathbf{C}^{*-1}\mathbf{P}+\mathbf{C} \right)^{-1}$ for $\mathbf{H}^{*}$
		\STATE \textbf{Return} $\mathbf{W}$, $\mathbf{b}$, $\boldsymbol{\alpha}$ and $\boldsymbol{\beta}$.
	\end{algorithmic}
	\label{algorithm:rcelmah}
\end{algorithm}
 The inherent channel effects in satellite communication result in the impropriety of received signals, which leads to non-zero pseudo-correlation. The estimation process can be significantly enhanced by considering and utilizing these additional second-order statistics. In this section, the weights calculation between the hidden and output layers is conducted by the widely linear regressor.

The formula of the regressor has been shown by (\ref{CELMAH_h2o}), where $\mathbf{H}$ and $\mathbf{H}^{*}$ are linearly processed by two regressors $\boldsymbol{\beta}$ and $\boldsymbol{\alpha}$ to reach the target $\mathbf{x}$. The specific formulas for widely linear regressors are derived as follows:
\begin{equation}
	\left( \mathbf{x}-\hat{\mathbf{x}} \right) \bot \ \mathbf{H},
	\label{equ:orthogonal1}
\end{equation}
and
\begin{equation}
	\left( \mathbf{x}-\hat{\mathbf{x}} \right) \bot \ \mathbf{H}^{*},
	\label{equ:orthogonal2}
\end{equation}
\noindent  where the symbol $\bot$ means that all the components of $\mathbf{H}$ and $\mathbf{H}^{*}$ are orthogonal to $\mathbf{x}-\hat{\mathbf{x}}$ with the scalar product. Therefore, (\ref{equ:orthogonal1}) can be expressed as:
\begin{equation}
	\begin{aligned}
		\mathbb{E} \left[ \left(\mathbf{x}-\hat{\mathbf{x}}\right)^H \mathbf{H} \right] &= 0 \\ \mathbb{E} \left[ \mathbf{x}^H \mathbf{H} \right] & = \mathbb{E} \left[ \boldsymbol{\beta}\mathbf{H}^{H}\mathbf{H} + \boldsymbol{\alpha}\mathbf{H}^{T}\mathbf{H} \right] ,
		\label{equ:wlp1}
	\end{aligned}
\end{equation}
\noindent and (\ref{equ:orthogonal2}) can be expressed as:
\begin{equation}
	\begin{aligned}
		\mathbb{E} \left[ \left(\mathbf{x}-\hat{\mathbf{x}}\right)^H \mathbf{H}^{*} \right] &= 0 \\ \mathbb{E} \left[ \mathbf{x}^H \mathbf{H}^{*} \right] & = \mathbb{E} \left[ \boldsymbol{\beta}\mathbf{H}^{H}\mathbf{H}^{*} + \boldsymbol{\alpha}\mathbf{H}^{T}\mathbf{H}^{*} \right] .
		\label{equ:wlp2}
	\end{aligned}
\end{equation}
To derive a practical solution, we replace the expectations with instantaneous values, enabling the formulation of a tractable closed-form estimator. Let instantaneous information replace the expectation, and we have $\mathbf{R}= \mathbf{x}^{H}\mathbf{H}, \mathbf{S}= \mathbf{x}^{T}\mathbf{H} ,\mathbf{P}= \mathbf{H}^{T}\mathbf{H} $ and $ \mathbf{C}= \mathbf{H}^{H}\mathbf{H} $. Then (\ref{equ:wlp1}) can be expressed as
\begin{equation}
	\mathbf{R}=\boldsymbol{\beta}\mathbf{C}+\boldsymbol{\alpha}\mathbf{P},
	\label{equ:wpl3}
\end{equation}
and simplified as
\begin{equation}
	\boldsymbol{\beta}=\left( \boldsymbol{\alpha}\mathbf{P} - \mathbf{R} \right)\mathbf{C}^{-1} .
	\label{equ:wpl3_1}
\end{equation}
This expression of $\boldsymbol{\beta}$ in terms of $\boldsymbol{\alpha}$ allows to eliminate $\boldsymbol{\beta}$ from the subsequent equations. And a similar derivation is applied to obtain the expression of $\boldsymbol{\alpha}$. Then, the (\ref{equ:wlp2}) can be expressed as
\begin{equation}
	\mathbf{S}^{*}=\boldsymbol{\beta}\mathbf{P}^{*}+\boldsymbol{\alpha}\mathbf{C}^{*} ,
	\label{equ:wpl4}
\end{equation}
\noindent and simplified to
\begin{equation}
	\boldsymbol{\alpha}=\left( \boldsymbol{\beta}\mathbf{P}^{*} - \mathbf{S}^{*}  \right)\mathbf{C}^{*-1} .
	\label{equ:wpl4_1}
\end{equation}
Finally, substituting (\ref{equ:wpl3_1}), the expression of $\boldsymbol{\beta}$, into the (\ref{equ:wpl4}) yields the closed-form estimator for $\boldsymbol{\alpha}$.
\begin{equation}
	\boldsymbol{\alpha} = \left( \mathbf{S}^{*}+\mathbf{R}\mathbf{C}^{-1}\mathbf{P}^{*} \right) \left( \mathbf{P}\mathbf{C}^{-1}\mathbf{P}^{*} + \mathbf{C}^{*} \right)^{-1}.
	\label{WLP_alpha}
\end{equation}
To obtain the expression of $\boldsymbol{\beta}$, similarly substitute (\ref{equ:wpl4_1}) into (\ref{equ:wpl3}), the regressor becomes
\begin{equation}
	\boldsymbol{\beta} = \left( \mathbf{R}+\mathbf{S}\mathbf{C}^{*-1}\mathbf{P} \right) \left( \mathbf{P}^{*}-\mathbf{C}^{*-1}\mathbf{P}+\mathbf{C} \right)^{-1}.
	\label{WLP_beta1}
\end{equation}
Therefore, the \gls{wlls} estimators for data impropriety design are derived, and the structure of this network is called CELM-WLLS

\subsection{Complexity Analysis}
In this section, we evaluate the complexity of different approaches based on three key metrics: complexity in terms of operating time, FLOPs (floating-point operations), and the overall algorithmic complexity $\mathcal{O}$. Operation time was measured using the MATLAB function \texttt{tictoc}, with all simulations conducted in a consistent environment using a \gls{dft-s-odfm} system where both the transmitted and received signal sizes are $1 \times 1024$. Since ML-based methods involve both channel estimation and equalization, the complexity measurements cover the entire process, from signal reception to symbol equalization. For fair comparisons, traditional methods were evaluated under the same conditions, with channel estimations relying on pilot signals. All simulations were performed on an Intel Core i7-4790 CPU running at 3.60 GHz, and the results are summarized in TABLE~\ref{tab:flops_time}.
\begin{table}[t]
	\centering
	\caption{FLOPs and Operation Time of Different Algorithms}
	\label{tab:flops_time}
	\begin{tabular}{@{}lcc@{}}
		\toprule
		\textbf{Algorithm} & \textbf{FLOPs} & \textbf{Operation Time (s)} \\ \midrule
		CELM      & 814768   & 0.0015  \\
		CELMAH    & 1791424  & 0.0028   \\
		CELMWLLS  & 558916  & 0.0011     \\
		LS        & 170011  & 2.9497 $\times 10^{-4}$  \\
		MMSE      & 196608  & 4.2021 $\times 10^{-4}$  \\ \bottomrule
	\end{tabular}
\end{table}
\subsubsection{Operation Time}
The results in TABLE~\ref{tab:flops_time} show that CELM-WLLS has the lowest execution time (0.0011 s), outperforming both CELMAH (0.0028 s) and CELM (0.0015 s). This is due to more efficient design of the CELM-WLLS algorithm, which reduces the number of required hidden neurons and optimizes the matrix inversion process. The \gls{ls} and \gls{mmse} methods, while having significantly lower computational requirements, are limited in their ability to handle complex satellite channel impairments, making them less robust in practice despite their faster computation times.
\subsubsection{FLOPs Analysis}
In terms of FLOPs, CELM-WLLS also exhibits superior performance, with a total of 558,916 FLOPs, compared to 1,791,424 for CELMAH and 814,768 for CELM. The lower FLOPs count in CELM-WLLS arises from its reliance on multiple small matrix operations rather than large matrix inversions. Specifically, while CELMAH requires the inversion of the augmented hidden layer matrix ${\Tilde{\boldsymbol{\beta}}}$, whose size depends on the length of the received signal and the number of hidden nodes, CELM-WLLS achieves similar or better performance by using smaller, more frequent matrix operations with reduced computational load.
\subsubsection{Algorithmic Complexity}
The complexity of CELMAH can be expressed as:
\begin{equation}
	\mathcal{O}\left( \mathrm{L} \times \mathrm{I} \times \mathrm{N} + \mathrm{N} \times (2\mathrm{L})^2  + (2\mathrm{L}) ^3 \right) ,
\end{equation}
where $\mathrm{L}$ is the number of hidden neurons, $\mathrm{I}$ is the number of input features, and $\mathrm{N}$ is the number of training samples. This expression captures the cost of computing the hidden layer output and performing matrix inversions. For CELM-WLLS, the complexity is slightly higher due to additional operations:
\begin{equation}
	\mathcal{O}\left( \mathrm{L} \times \mathrm{I} \times \mathrm{N} + \mathrm{N} \times \mathrm{L}^2 + \mathrm{L}^3 \right) ,
\end{equation}
where the term $\mathrm{L}^3$ accounts for the inversion of smaller matrices multiple times. Although the complexity of CELM-WLLS includes an extra cubic term, the reduced size of the matrices being inverted results in lower overall FLOPs, as shown in the table.

Despite the higher theoretical complexity of CELM-WLLS, its lower FLOPs and faster execution time make it a more efficient solution compared to CELMAH. The efficiency gains in CELM-WLLS stem from using smaller, and more frequent matrix operations, which significantly reduces computational requirements without compromising accuracy. This advantage is reflected in both the operation time and FLOPs analysis, where CELM-WLLS outperforms CELMAH while maintaining robust performance in satellite communication scenarios.

%%%%%%%%%%%%%%%%%%%%%%%%%%%%%%%%%%%%%%%%%%%%%%%%%%%%%%%%%%%%%%%%%%%%%%%
\section{Simulation Results} \label{sec04}
\begin{figure*}[!t]
	\centering
	\subfigure[Pilot estimation comparison of LS and ML approaches]{
		\includegraphics[width=\mysize]{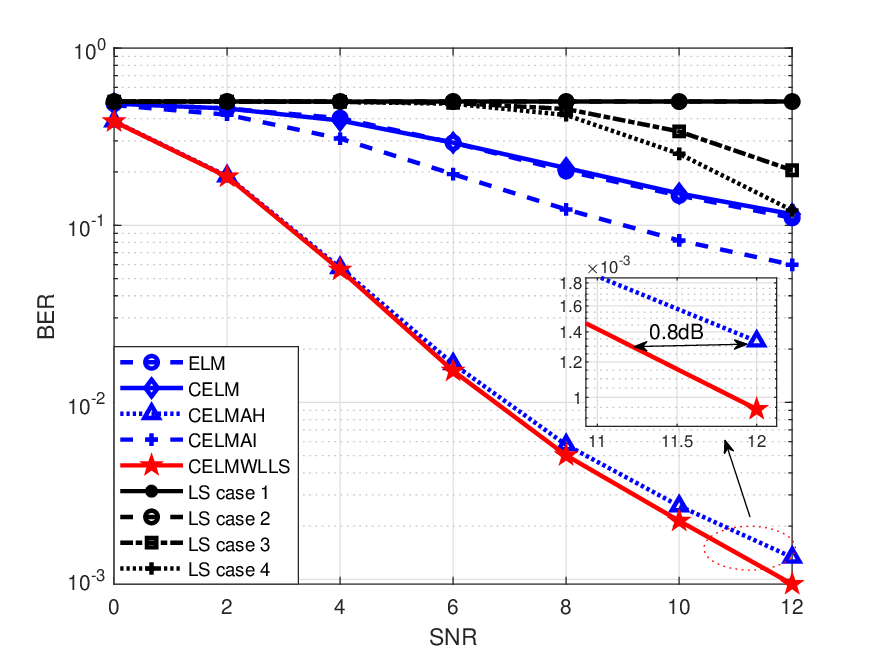}
		\label{fig0_pilot_LS}
	}
	\quad
	\subfigure[Pilot estimation comparison of MMSE and ML approaches]{
		\includegraphics[width=\mysize]{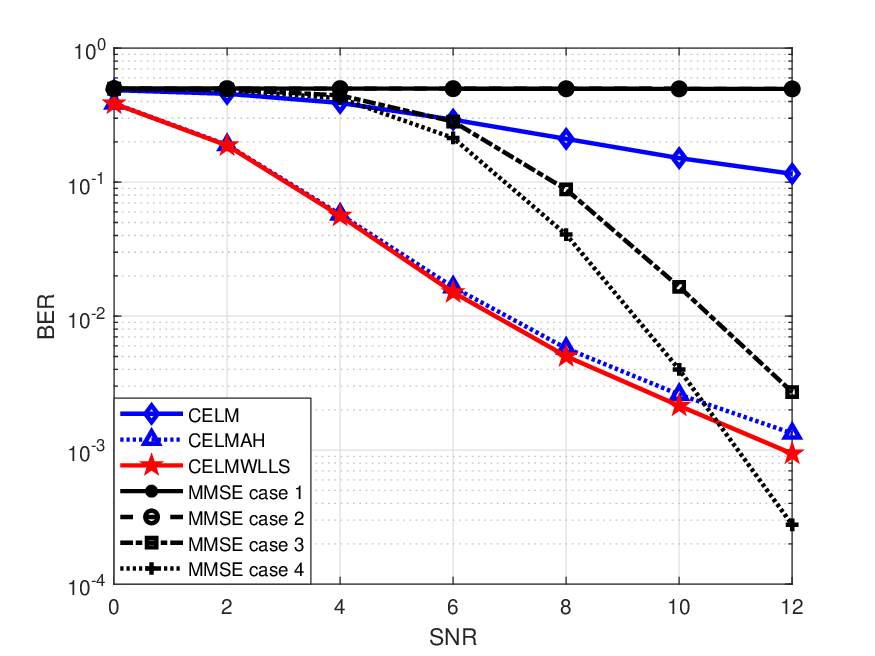}
		\label{fig0_pilot_MMSE}
	}
	\caption{BER versus SNR comparison of ML-based and traditional methods with pilot channel estimation in maximum residual Doppler shifts of $f_d = 1000$ Hz,  multi-path delay spread of $10$ ns, I/Q phase imbalance of $1.39$ rad and without IBO}
	\label{fig0_pilot}
\end{figure*}
This section presents the simulation results to demonstrate the effectiveness of the CELM-WLLS based approache under satellite-to-ground-station channel impairments. The scenario is considered as a LEO satellite downlink to ground station with a tangential velocity of 7.61 km/s, an altitude of 800 km, and an S-band carrier frequency of 2 GHz. Due to the high relative velocity between the satellite and user equipment, the maximum Doppler shift can reach up to 44.1 kHz, significantly exceeding the frequency error tolerance defined by NR standards (5 \gls{ppm}, or 10 kHz for the S-band). Although current 5G specifications do not account for such high Doppler shifts, the use of satellite ephemeris data and the user’s location allows for the application of blind coarse frequency shift correction, reducing the residual Doppler shift within the 5 \gls{ppm} limit \cite{jayaprakash2020new}. The AM/AM and AM/PM characteristics of a typical satellite TWTA, provided by the European Space Agency \cite{jayaprakash2019analysis}, represent the non-linearity of the satellite. The multi-path delay is maximized at 100 ns, which is the nominal delay spread for \gls{nlos} conditions \cite{huang2021synchronization}. The air interface waveform used is a fully loaded \gls{dft-s-odfm} waveform with a \gls{fft} size of $N = 1024$, sub-carrier spacing of $\Delta f = 15$ kHz, and a cyclic prefix length of $N_{CP} = 72$, modulated using 4-\gls{qam}. Given the structure of DFT-s-OFDM introduces spectral correlation across subcarriers, a channel estimate obtained from a single DFT-s-OFDM symbol can be effectively applied to other symbols within the same transmission frame. Therefore, in this work, a dedicated DFT-s-OFDM symbol serves as the pilot block for channel estimation, providing the CSI for subsequent processing. This estimated CSI is then employed for detection in the subsequent data-bearing DFT-s-OFDM symbols. For the online training of the network, the input weights and biases follow a uniform distribution with zero mean and variance $\theta$. The activation function used in the hidden layer is the complex inverse hyperbolic sine function. To balance the BER performance and network complexity, the number of hidden nodes was gradually increased, and the optimal number of hidden nodes $N_h$ for both CELM and CELMAH was determined to be 6 using cross-validation. Additionally, all simulations are conducted under the following default parameters: a maximum residual Doppler shift of $f_d = 1000$ Hz, a delay spread of $10$ ns, an I/Q phase imbalance of $1.39$ rad, and without IBO. Under these default settings, the normalized Doppler shift is $f_d / \Delta f = 1000 / 15000 = 6.6\%$, indicating a minor Doppler effect relative to the subcarrier spacing. Since a typical DFT-s-OFDM frame lasts around 1~ms, which is much shorter than the channel coherence time, the channel can be considered quasi-static within each frame. This ensures consistent statistical properties between training and inference data, effectively avoiding the out-of-distribution issue common in time-varying or mismatched channels.
\begin{figure*}[!t]
	\centering
	\subfigure[Estimation comparison of LS with CSI vs. pilot based ML approaches]{
		\includegraphics[width=\mysize]{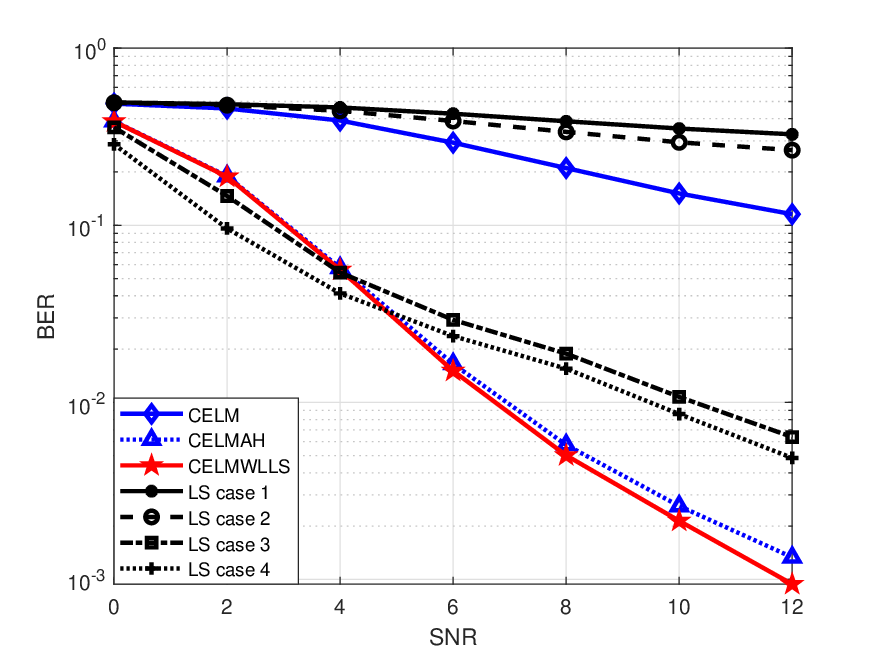}
		\label{fig0_LS}
	}
	\quad
	\subfigure[Estimation comparison of MMSE with CSI vs. pilot based ML approaches]{
		\includegraphics[width=\mysize]{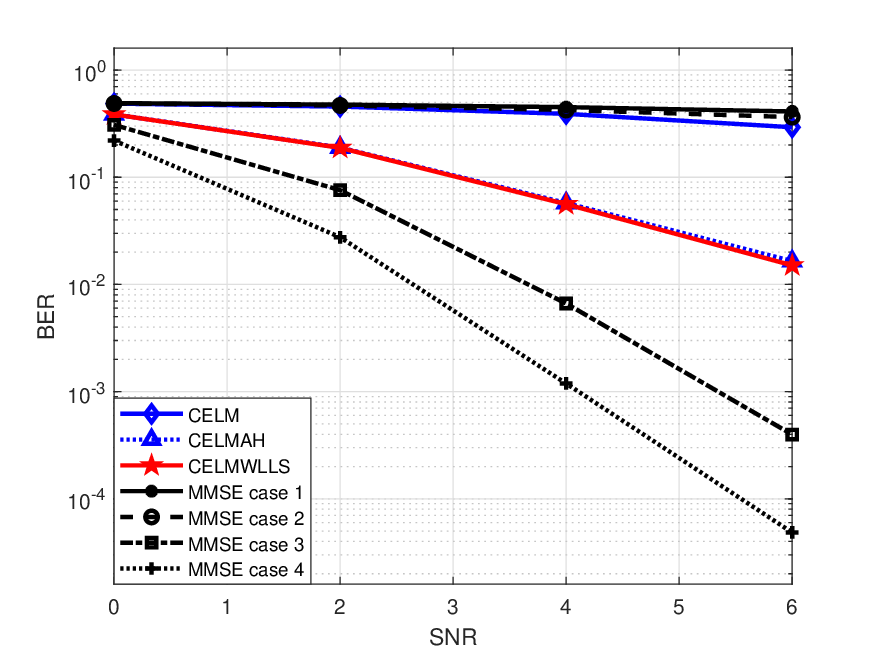}
		\label{fig0_MMSE}
	}
	\caption{BER versus SNR comparison where ML-based methods rely on pilot channel estimation while traditional methods assume perfect CSI.}
	\label{fig0_full_CSI}
\end{figure*}

Fig.~\ref{fig0_pilot} presents a comparison of the \gls{ber} versus \gls{snr} performance for conventional approaches (LS and MMSE) and ML-based approaches (ELM, CELM, CELMAH, and CELM-WLLS) under pilot-based channel estimation. The results highlight two critical challenges in satellite communication: non-linear distortion and signal impropriety. To assess the performance of ML-based methods, there are no HPA pre-distortion or impairment compensation applied in their simulations. In contrast, the traditional methods (LS and MMSE) are evaluated under four progressively improved conditions:
\begin{itemize}
	\item[$\bullet$] case 1: without HPA pre-distortion and exist all impairments
	\item[$\bullet$] case 2: with ideal HPA pre-distortion but exist all impairments
	\item[$\bullet$] case 3: with ideal HPA pre-distortion and without impropriety (other impairments exist)
	\item[$\bullet$] case 4: with ideal HPA pre-distortion and without any impairment
\end{itemize}
For the traditional methods, to mitigate the non-linearity introduced by the HPA, a pre-distortion technique can be employed to linearize the power amplifier. However, practical pre-distortion methods generally provide only partial compensation, resulting in residual distortions. Since this study primarily focuses on evaluating machine learning-based approaches, traditional methods are assumed to incorporate an ideal pre-distorter as a benchmark. Specifically, the ideal HPA pre-distorter ensures that the amplitude and phase characteristics of the HPA output align with those of an ideal linear amplifier, thereby entirely eliminating AM/AM and AM/PM distortions \cite{silva2005survey}. Furthermore, the impairments considered in this work include all degradation factors illustrated in Fig.~\ref{Satellite_Channel}. Among these, phase noise and I/Q imbalance refer to signal impropriety.

Fig.~\ref{fig0_pilot_LS} and Fig.~\ref{fig0_pilot_MMSE} show the performance comparison between CELM based approaches with the traditional LS and MMSE methods, respectively. Apparently, conventional methods suffer significantly from the non-linearity and impropriety of the satellite channel, severely impacting pilot channel estimation and equalization. Even with pre-distortion, impropriety still has a severe effect, as indicated by case 2, where signal recovery is essentially unsuccessful. The performance of case 3 illustrates the BER performance improvement of LS and MMSE in the absence of impropriety. The case 4 performance curve of an ideal pre-distorter without satellite channel impairments is also included to facilitate comparison. Notably, this work set the state of impropriety in case 3 and the state of impairments in case 4 were eliminated under ideal compensation. Although this is unfair compared with the ML series of methods, cases 3 and 4 could intuitively reflect the impact of impropriety or impairments on performance as the benchmarks. For completeness, the performance of the traditional real-valued ELM is also included. While ELM is a universal approximator in the real domain and thus can partially model impairments, it lacks structural modeling for signal impropriety. Due to the shared network configuration and the absence of a widely linear design, its BER performance nearly overlaps with the baseline CELM. Additionally, the other scheme proposed in \cite{zhang2018augmented}, referred to as \gls{celmai}, is shown in Fig.~\ref{fig0_pilot_LS}, and the results indicate that the input layer augmented structure is unsuitable for current communication scenarios. These results also confirm the theoretical analysis that the additional second-order statistics attributed to the impropriety of the received signal are utilized in both CELMAH and CELM-WLLS. Additionally, the estimator of CELM-WLLS is designed for signal impropriety, and this variant shows approximately a gain of 0.8 dB over CELMAH. The observed performance gain primarily originates from the weight computation strategy of CELM-WLLS, which more rigorously adheres to the principles of WLP. Specifically, CELMAH, as proposed in \cite{zhang2018augmented}, employs a direct least squares solution to compute the weight matrix, as given in (\ref{lesat_square}). In contrast, CELM-WLLS, introduced in this work, refines this approach by independently estimating the weights for the hidden layer and the augmented hidden layer nodes, leveraging the widely linear regression formulation in (\ref{WLP_alpha}) and (\ref{WLP_beta1}). This structural distinction enables CELM-WLLS to more effectively exploit the statistical properties of improper signals, thereby improving estimation accuracy and enhancing robustness in satellite communication scenarios.

As a baseline comparison, Fig.~\ref{fig0_full_CSI} presents the BER versus SNR performance of the CELM series methods with pilot channel estimation, compared to traditional approaches with full CSI knowledge. The simulation parameters are the same as the default. As shown in Fig.~\ref{fig0_LS} and \ref{fig0_MMSE}, non-linearity and impropriety still significantly impact traditional methods. Due to algorithmic factors, the LS approach achieves gains from the known channel fading but cannot outperform CELMAH and CELM-WLLS at high SNRs. Correspondingly, the MMSE algorithm exhibits considerable improvements with the condition of knowing CSI, as shown in Fig.~\ref{fig0_MMSE}. The results of cases 3 and 4 are obtained by removing the most significant satellite channel impairments (non-linearity, impropriety and impairments), they serve as lower bounds.
\begin{figure*}[htbp]
	\begin{minipage}[t]{0.5\linewidth}
		\includegraphics[width=\mysize]{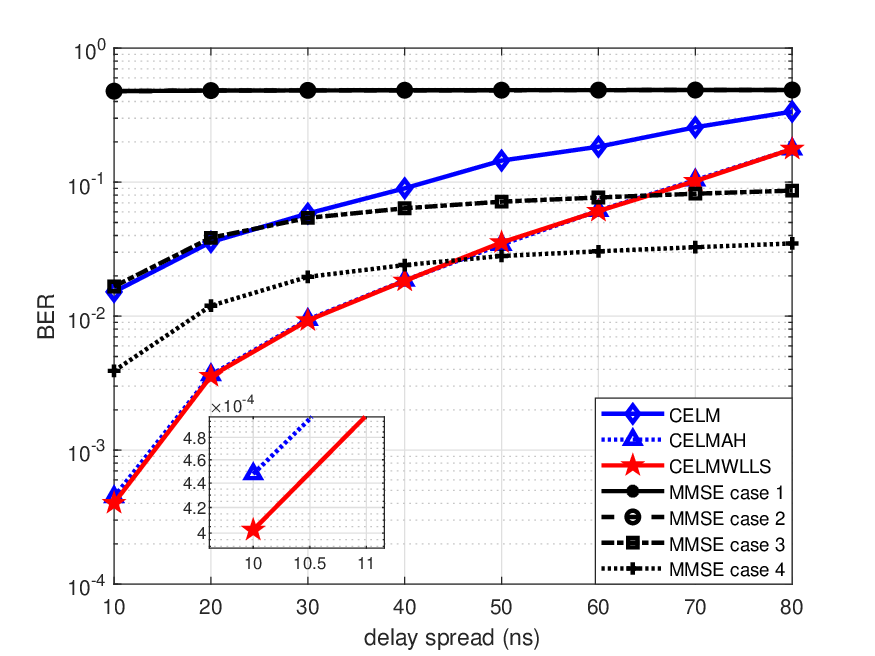}
		\caption{BER versus multi-path channel delay spread comparison}
		\label{fig_delayspread_1}
	\end{minipage}%
	\begin{minipage}[t]{0.5\linewidth}
		\includegraphics[width=\mysize]{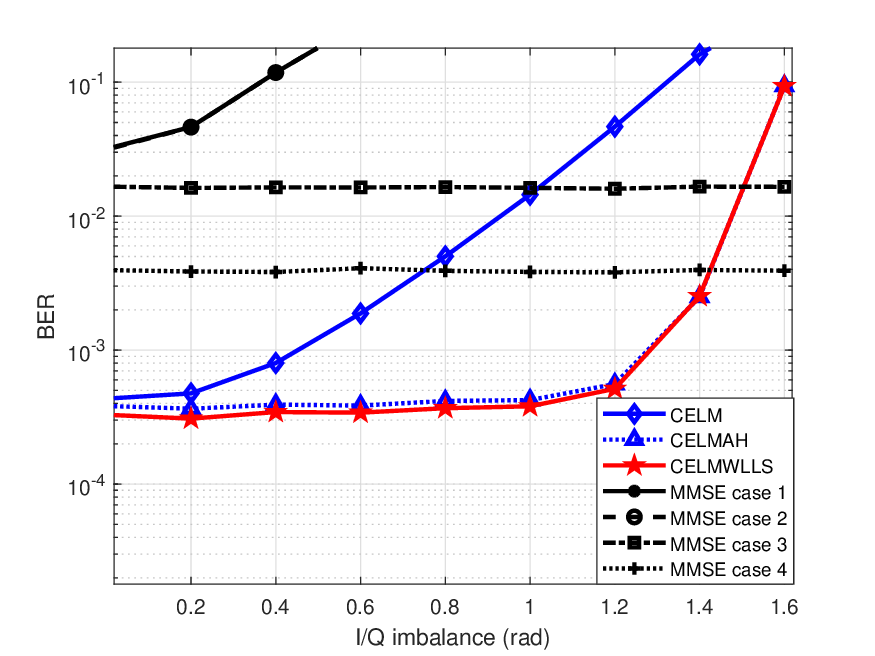}
		\caption{BER performance comparison in different IQ phase imbalance}
		\label{fig_iqi_1}
		\vspace{-1em}
	\end{minipage}
\end{figure*}
\begin{figure*}[!t]
	\begin{minipage}[t]{0.5\linewidth}
		\centering
		\includegraphics[width=\mysize]{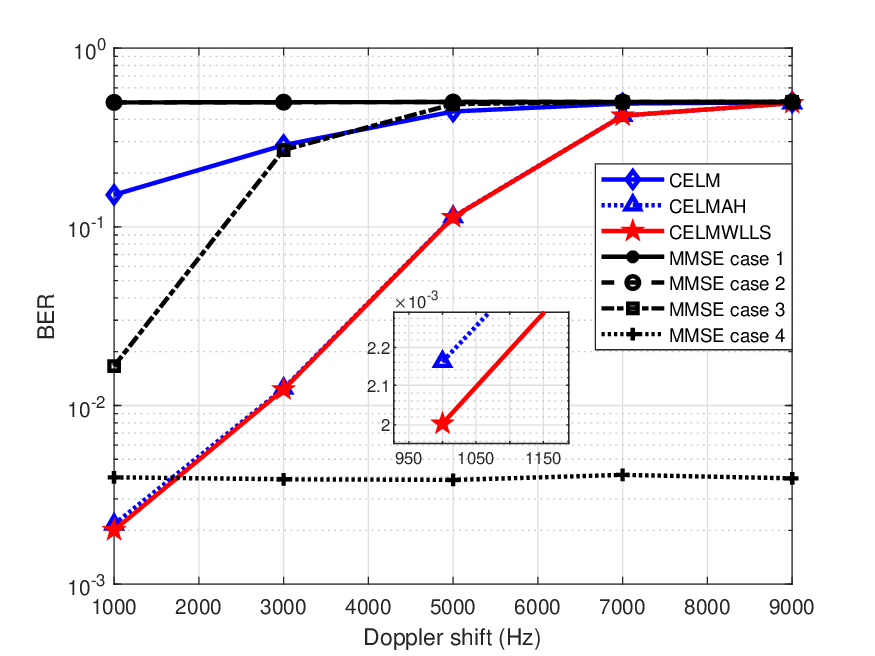}
		\caption{BER versus Doppler shift performance comparison in 10 dB SNR}
		\label{fig_doppler_fixsnr}
	\end{minipage}
	\begin{minipage}[t]{0.5\linewidth}
		\centering
		\includegraphics[scale=0.56]{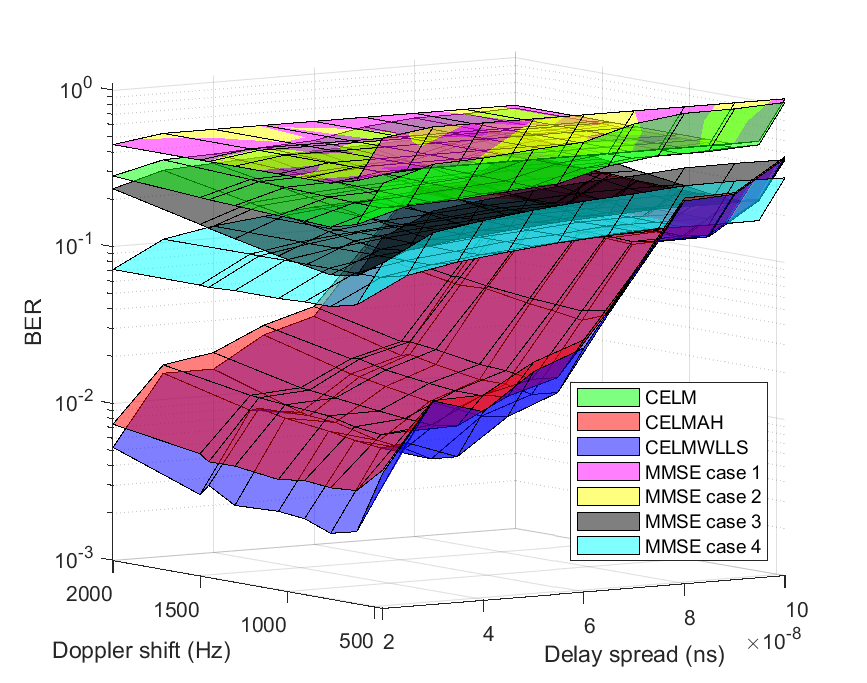}
		\caption{BER versus multi-path delay spread and residual Doppler shift comparison for coded system at S-band}
		\label{fig_dsds}
	\end{minipage}%
\end{figure*}

The comparisons above demonstrate that MMSE consistently outperforms LS in traditional methods. Therefore, to simplify the comparative results, subsequent simulations will focus on comparing MMSE with the ML-based methods. Additionally, to ensure fairness, the CSI for all methods will be derived and utilized through the pilot signal channel estimation techniques in the simulations below.

The BER performance comparisons for multi-path channel delay spread are shown in Fig.~\ref{fig_delayspread_1}, other parameters are the same as the default. Fig.~\ref{fig_delayspread_1} illustrates the impact of varying delay spreads on the BER performance of different approaches at an SNR of 10 dB. Observing case 1, we can find that MMSE fails to operate over satellite channels. Even case 2 with ideal HPA pre-distortion still faces challenges. Only when the impropriety or all impairments are removed can it outperform CELMAH or CELM-WLLS in high-delay conditions. Notably, in the simulation of Fig.~\ref{fig_delayspread_1}, the number of neurons in the hidden layer of the networks is 6, which is set to a small number to reduce the complexity of online learning. When dealing with complex channels, the number of neurons can be adjusted to adapt to different conditions.

To assess the robustness of CELM-WLLS in mitigating the effects of signal impropriety in the received constellation, Fig.~\ref{fig_iqi_1} presents a comparative analysis of ML-based methods and MMSE. Specifically, the figure illustrates the BER performance as a function of varying I/Q phase imbalance levels in a coded system operating at the S-band, with an SNR of 10 dB. As the level of impropriety increases, CELM exhibits a noticeable degradation in BER performance. In contrast, CELMAH and CELM-WLLS initially maintain relatively stable BER; however, once the I/Q imbalance surpasses a certain threshold, the performance of both models deteriorates due to the limitations of their current network parameterization. Additionally, the results for MMSE under cases 1 and 2 exhibit significant overlap, indicating a failure in effective signal recovery under these conditions. In contrast, the performance remains unchanged in cases 3 and 4, as I/Q phase imbalance is not considered in these scenarios. These results further highlight that CELM-WLLS demonstrates robustness against I/Q imbalance compared to other approaches.
\begin{figure}[t!]
	\centering
	\includegraphics[scale=0.65]{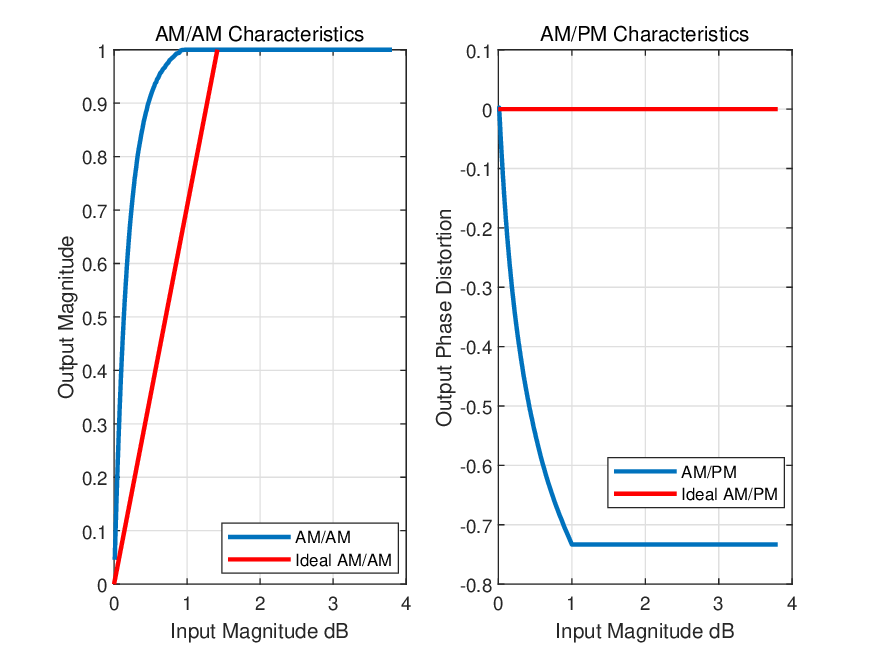}
	\caption{AM/AM and AM/PM characteristics of the TWTA HAP}
	\label{fig_hpa_twta}
\end{figure}

Doppler shift is a critical challenge in the non-geostationary orbit satellite domain, and its impact is demonstrated in Fig.~\ref{fig_doppler_fixsnr}, which illustrates the BER performance across varying Doppler shifts at an SNR of 10 dB. The results clearly indicate that both CELMAH and the proposed CELM-WLLS exhibit the most robust performance, with the latter demonstrating a slight advantage. Notably, the proposed method maintains a performance gain over the former, corresponding to approximately a 100 Hz Doppler shift gap, highlighting its enhanced resilience to Doppler-induced distortions. In contrast, CELM is significantly more susceptible to performance degradation as the Doppler shift increases. Additionally, MMSE case 4 maintains the best and most stable performance, as it serves as the ideal benchmark and remains unaffected by the Doppler shift.

To demonstrate the capability of CELMAH in addressing various impairments across different delay spreads and Doppler shift scenarios, we compare its performance under an SNR of 8 dB, as shown in Fig.~\ref{fig_dsds}. The results indicate that CELM-WLLS outperforms all other schemes, with CELMAH approaching, but not surpassing, CELM-WLLS in performance. Standard CELM performs better than MMSE in both cases 1 and 2, suggesting that while CELM offers some compensation for non-linearity, it is less effective against satellite channel impairments. The performance comparison highlights how CELM-WLLS and CELMAH effectively leverage signal impropriety caused by satellite impairments, resulting in superior overall performance. However, MMSE in cases 1 and 2 fails to deliver accurate channel estimation and equalization, leading to overlapping and inconsistent results. Furthermore, even when impairments are excluded in case 4, the presence of aggravated \gls{isi} and multi-path fading still degrade performance, as evidenced when compared to CELM-based approaches. The overall performance in Fig.~\ref{fig_dsds} underscores the limitations of MMSE in handling satellite channel impairments, while CELMAH demonstrates robust performance across varying delay spreads and Doppler shifts.

\begin{figure}[!t]
	\centering
	\subfigure[BER versus SNR comparison with 4-QAM]{
		\includegraphics[width=\mysize]{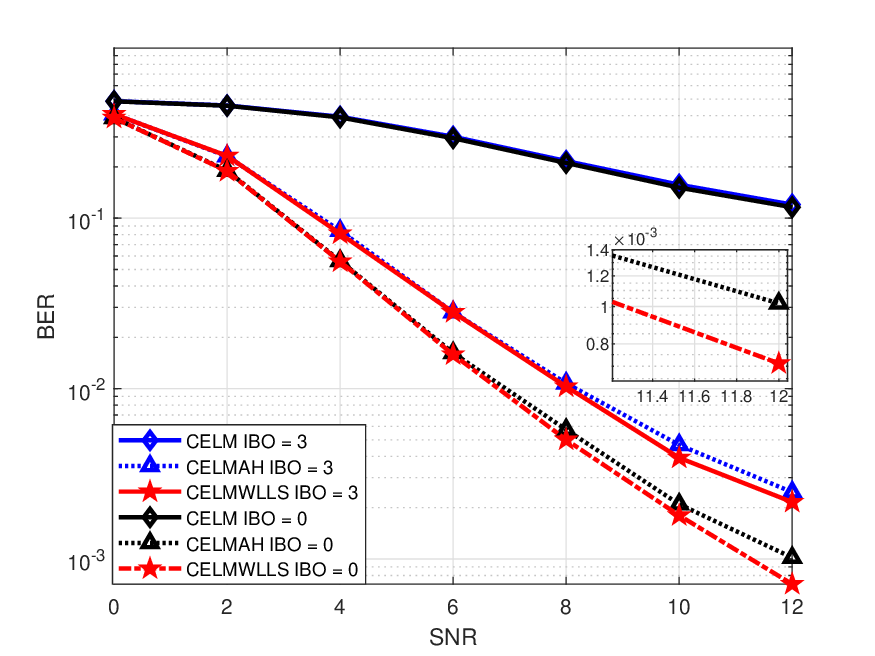}
		\label{fig_ibo_4qam}
	}
	\quad
	\subfigure[BER versus IBO comparison with 16-QAM]{
		\includegraphics[width=\mysize]{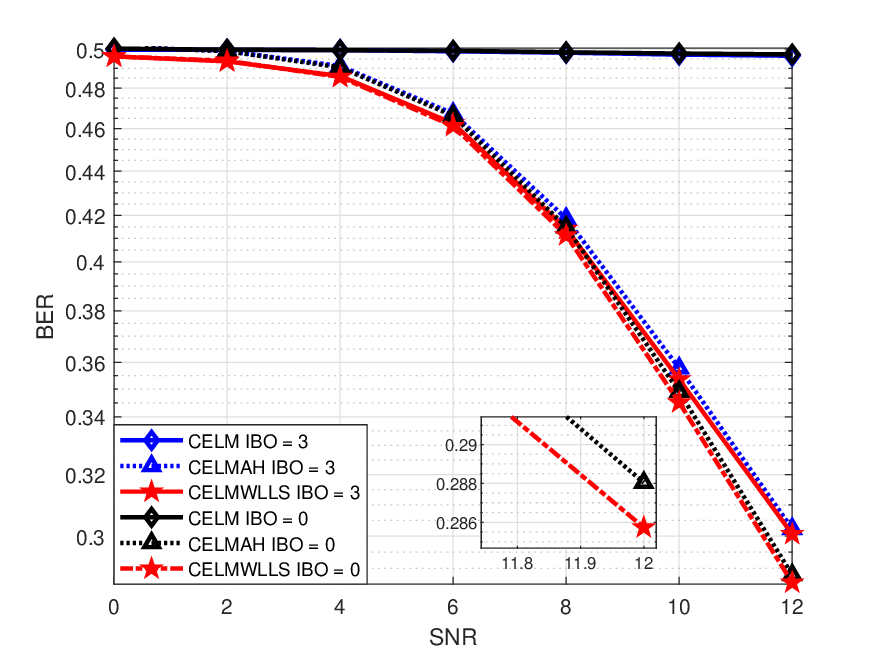}
		\label{fig_ibo_16qam}
	}
	\caption{BER performance comparison for IBO in 4-QAM and 16-QAM}
	\label{fig_ibo}
\end{figure}
The non-linearity of the HPA poses a significant challenge in satellite communication systems, which is mitigated in this work by the ML models' ability to generalize non-linear effects. As illustrated in Fig.~\ref{fig_hpa_twta}, the AM/AM and AM/PM characteristics of the TWTA HPA show typical non-linear behavior, including clipping at high power levels that affect both amplitude and phase. To minimize these distortions, IBO is typically applied to reduce input power, ensuring that the HPA operates in a region with lower non-linearity. However, excessively high IBO values reduce average transmitted power, leading to a performance trade-off in traditional methods. While MMSE-based approaches (cases 2, 3, and 4) assume ideal HPA pre-distortion, masking the effect of IBO, case 1 cannot compensate for satellite channel impairments, limiting its effectiveness compared to ML-based approaches, as shown in Fig.~\ref{fig_ibo}. Specifically, Fig.~\ref{fig_ibo_4qam} shows the simulation results of the 4-QAM modulation under IBO values in 0 dB and 3 dB. The results show that CELM-WLLS consistently outperforms CELMAH, while CELM improves with increasing SNR but falls short in terms of accuracy. Increasing IBO negatively impacts BER, demonstrating that both CELMAH and the proposed method effectively compensate for non-linearities introduced by the HPA. However, applying unnecessary IBO reduces transmitted power, resulting in degraded performance. This effect is more pronounced with higher-order modulation, as evidenced by the significant BER gap under 16-QAM shown in Fig.~\ref{fig_ibo_16qam}. Ultimately, CELMAH and CELM-WLLS maintain high transmission accuracy without requiring IBO, making them more resource-efficient for satellite transmitters.
%%%%%%%%%%%%%%%%%%%%%%%%%%%%%%%%%%%%%%%%%%%%%%%%%%%%%%%%%%%%%%%%%%%%%%%
\section{Conclusion} \label{sec05}
This paper presented the design and analysis of CELM-based post-distortion schemes to improve satellite communication performance. By addressing the inherent impairments in satellite channels, we developed a novel approach that leverages the properties of improper signals to extract additional second-order statistics. The structural characteristics of the CELMAH network were used to replace traditional equalization and demodulation methods, while a tailored WL-LS formulation was derived to exploit widely linear processing. This led to the introduction of the CELM-WLLS approach, which enhances communication robustness and is particularly effective in dynamic satellite communication environments characterized by non-linear impairments. Simulation results demonstrated that CELM-WLLS not only achieved an approximate 0.8 dB gain in BER performance compared to CELMAH but also reduced computational complexity by approximately two-thirds. This improvement in both performance and efficiency makes CELM-WLLS a more practical and viable solution for satellite communications, especially when compared to traditional methods like MMSE. While traditional techniques may still show potential with ideal pre-distortion and full impairment compensation, the additional complexity of these methods underscores the advantage of ML-based approaches in reducing the computational burden. The proposed CELM-based schemes, particularly CELM-WLLS, demonstrate a well-balanced solution, offering significant performance enhancements alongside optimized computational efficiency.

\ifCLASSOPTIONcaptionsoff
\newpage
\fi

\bibliographystyle{IEEEtran}
\bibliography{IEEEabrv,biblist}

\begin{thebibliography}{10}
\providecommand{\url}[1]{#1}
\csname url@samestyle\endcsname
\providecommand{\newblock}{\relax}
\providecommand{\bibinfo}[2]{#2}
\providecommand{\BIBentrySTDinterwordspacing}{\spaceskip=0pt\relax}
\providecommand{\BIBentryALTinterwordstretchfactor}{4}
\providecommand{\BIBentryALTinterwordspacing}{\spaceskip=\fontdimen2\font plus
\BIBentryALTinterwordstretchfactor\fontdimen3\font minus
  \fontdimen4\font\relax}
\providecommand{\BIBforeignlanguage}[2]{{%
\expandafter\ifx\csname l@#1\endcsname\relax
\typeout{** WARNING: IEEEtran.bst: No hyphenation pattern has been}%
\typeout{** loaded for the language `#1'. Using the pattern for}%
\typeout{** the default language instead.}%
\else
\language=\csname l@#1\endcsname
\fi
#2}}
\providecommand{\BIBdecl}{\relax}
\BIBdecl

\bibitem{jiang2023software}
W.~Jiang, ``Software defined satellite networks: A survey,'' \emph{Digital
  Communications and Networks}, vol.~9, no.~6, pp. 1243--1264, 2023.

\bibitem{10179219}
J.~Heo, S.~Sung, H.~Lee, I.~Hwang, and D.~Hong, ``{M}{I}{M}{O} satellite
  communication systems: A survey from the phy layer perspective,'' \emph{IEEE
  Communications Surveys and Tutorials}, vol.~25, no.~3, pp. 1543--1570, 2023.

\bibitem{guidotti2019architectures}
A.~Guidotti, A.~Vanelli-Coralli, M.~Conti, S.~Andrenacci, S.~Chatzinotas,
  N.~Maturo, B.~Evans, A.~Awoseyila, A.~Ugolini, T.~Foggi \emph{et~al.},
  ``Architectures and key technical challenges for 5{G} systems incorporating
  satellites,'' \emph{IEEE Transactions on Vehicular Technology}, vol.~68,
  no.~3, pp. 2624--2639, 2019.

\bibitem{huang2024joint}
C.~Huang, G.~Chen, P.~Xiao, Y.~Xiao, Z.~Han, and J.~A. Chambers, ``Joint
  offloading and resource allocation for hybrid cloud and edge computing in
  {SAGIN}s: A decision assisted hybrid action space deep reinforcement learning
  approach,'' \emph{IEEE Journal on Selected Areas in Communications}, vol.~42,
  no.~5, pp. 1029--1043, May. 2024.

\bibitem{9792424}
Q.~Chen, Y.~Li, F.~Jalili, Z.~Wang, O.~Kiel~Jensen, G.~Frølund~Pedersen, and
  M.~Shen, ``Digital signal recovery with transmitter nonlinear state tracking
  for satellite communications,'' \emph{IEEE Transactions on Circuits and
  Systems II: Express Briefs}, vol.~69, no.~12, pp. 4774--4778, 2022.

\bibitem{10246293}
L.~Xu, J.~Jiao, Y.~Wang, S.~Wu, R.~Lu, and Q.~Zhang, ``Low-correlation
  superimposed pilot grant-free massive access for satellite internet of
  things,'' \emph{IEEE Transactions on Communications}, vol.~71, no.~12, pp.
  7087--7101, 2023.

\bibitem{10066300}
K.-X. Li, X.~Gao, and X.-G. Xia, ``Channel estimation for {L}{E}{O} satellite
  massive {M}{I}{M}{O} {O}{F}{D}{M} communications,'' \emph{IEEE Transactions
  on Wireless Communications}, vol.~22, no.~11, pp. 7537--7550, 2023.

\bibitem{9944375}
J.-H. Lee, J.~Park, M.~Bennis, and Y.-C. Ko, ``Integrating {L}{E}{O} satellites
  and multi-uav reinforcement learning for hybrid fso/rf non-terrestrial
  networks,'' \emph{IEEE Transactions on Vehicular Technology}, vol.~72, no.~3,
  pp. 3647--3662, 2023.

\bibitem{10273640}
Z.~Yan and D.~Li, ``Convergence time optimization for decentralized federated
  learning with {L}{E}{O} satellites via number control,'' \emph{IEEE
  Transactions on Vehicular Technology}, vol.~73, no.~3, pp. 4517--4522, 2024.

\bibitem{9999531}
S.~Song, J.~Liu, J.~Guo, B.~Lin, Q.~Ye, and J.~Cui, ``Efficient data collection
  scheme for multi-modal underwater sensor networks based on deep reinforcement
  learning,'' \emph{IEEE Transactions on Vehicular Technology}, vol.~72, no.~5,
  pp. 6558--6570, 2023.

\bibitem{10328395}
S.~Zhang, T.~Cai, D.~Wu, D.~Schupke, N.~Ansari, and C.~Cavdar, ``Iort data
  collection with {L}{E}{O} satellite-assisted and cache-enabled uav: A deep
  reinforcement learning approach,'' \emph{IEEE Transactions on Vehicular
  Technology}, vol.~73, no.~4, pp. 5872--5884, 2024.

\bibitem{algan2021image}
G.~Algan and I.~Ulusoy, ``Image classification with deep learning in the
  presence of noisy labels: A survey,'' \emph{Knowledge-Based Systems}, vol.
  215, p. 106771, 2021.

\bibitem{9546673}
J.~Liu, T.~Oyedare, and J.-M. Park, ``Detecting out-of-distribution data in
  wireless communications applications of deep learning,'' \emph{IEEE
  Transactions on Wireless Communications}, vol.~21, no.~4, pp. 2476--2487,
  2022.

\bibitem{9310357}
D.~Gao, Q.~Guo, and Y.~C. Eldar, ``Massive {M}{I}{M}{O} as an extreme learning
  machine,'' \emph{IEEE Transactions on Vehicular Technology}, vol.~70, no.~1,
  pp. 1046--1050, 2021.

\bibitem{6125249}
R.~Zhang, Y.~Lan, G.-B. Huang, and Z.-B. Xu, ``Universal approximation of
  extreme learning machine with adaptive growth of hidden nodes,'' \emph{IEEE
  Transactions on Neural Networks and Learning Systems}, vol.~23, no.~2, pp.
  365--371, 2012.

\bibitem{csahin2014application}
M.~{\c{S}}ahin, Y.~Kaya, M.~Uyar, and S.~Y{\i}ld{\i}r{\i}m, ``Application of
  extreme learning machine for estimating solar radiation from satellite
  data,'' \emph{International Journal of Energy Research}, vol.~38, no.~2, pp.
  205--212, 2014.

\bibitem{8715649}
J.~Liu, K.~Mei, X.~Zhang, D.~Ma, and J.~Wei, ``Online extreme learning
  machine-based channel estimation and equalization for {O}{F}{D}{M} systems,''
  \emph{IEEE Communications Letters}, vol.~23, no.~7, pp. 1276--1279, 2019.

\bibitem{soares2016adaptive}
S.~G. Soares and R.~Ara{\'u}jo, ``An adaptive ensemble of on-line extreme
  learning machines with variable forgetting factor for dynamic system
  prediction,'' \emph{Neurocomputing}, vol. 171, pp. 693--707, 2016.

\bibitem{javed2015new}
K.~Javed, R.~Gouriveau, and N.~Zerhouni, ``A new multivariate approach for
  prognostics based on extreme learning machine and fuzzy clustering,''
  \emph{IEEE Transactions on Cybernetics}, vol.~45, no.~12, pp. 2626--2639,
  2015.

\bibitem{zhang2014domain}
L.~Zhang and D.~Zhang, ``Domain adaptation extreme learning machines for drift
  compensation in e-nose systems,'' \emph{IEEE Transactions on instrumentation
  and measurement}, vol.~64, no.~7, pp. 1790--1801, 2014.

\bibitem{chen2018domain}
Y.~Chen, S.~Song, S.~Li, L.~Yang, and C.~Wu, ``Domain space transfer extreme
  learning machine for domain adaptation,'' \emph{IEEE Transactions on
  Cybernetics}, vol.~49, no.~5, pp. 1909--1922, 2018.

\bibitem{9505614}
D.~Wang, P.~Qi, Y.~Zhao, C.~Li, W.~Wu, and Z.~Li, ``Covert wireless
  communication with noise uncertainty in space-air-ground integrated vehicular
  networks,'' \emph{IEEE Transactions on Intelligent Transportation Systems},
  vol.~23, no.~3, pp. 2784--2797, 2022.

\bibitem{5961645}
T.~Adali, P.~J. Schreier, and L.~L. Scharf, ``Complex-valued signal processing:
  The proper way to deal with impropriety,'' \emph{IEEE Transactions on Signal
  Processing}, vol.~59, no.~11, pp. 5101--5125, 2011.

\bibitem{chen2021widely}
Y.~Chen, L.~You, A.-A. Lu, and X.~Gao, ``Widely-linear processing for the
  uplink of the massive {M}{I}{M}{O} with {I}{Q} imbalance: Channel estimation
  and data detection,'' \emph{IEEE Transactions on Signal Processing}, vol.~69,
  pp. 4685--4698, 2021.

\bibitem{zhang2021adaptive}
S.~Zhang, J.~Zhang, Y.~Xia, and H.~C. So, ``Adaptive frequency-domain
  normalized implementations of widely-linear complex-valued filter,''
  \emph{IEEE Transactions on Signal Processing}, vol.~69, pp. 5801--5814, 2021.

\bibitem{zhang2018augmented}
H.~Zhang, Y.~Wang, D.~Xu, J.~Wang, and L.~Xu, ``The augmented complex-valued
  extreme learning machine,'' \emph{Neurocomputing}, vol. 311, pp. 363--372,
  2018.

\bibitem{jayaprakash2020new}
A.~Jayaprakash, B.~G. Evans, P.~Xiao, A.~B. Awoseyila, and Y.~Zhang, ``New
  radio numerology and waveform evaluation for satellite integration into 5{G}
  terrestrial network,'' in \emph{ICC 2020-2020 IEEE International Conference
  on Communications (ICC)}.\hskip 1em plus 0.5em minus 0.4em\relax IEEE, 2020,
  pp. 1--7.

\bibitem{o2009new}
M.~O'droma, S.~Meza, and Y.~Lei, ``New modified saleh models for memoryless
  nonlinear power amplifier behavioural modelling,'' \emph{IEEE Communications
  Letters}, vol.~13, no.~6, pp. 399--401, 2009.

\bibitem{3gpp2019study}
{3GPP}, ``{Study on New Radio (NR) to support non-terrestrial networks},'' {3rd
  Generation Partnership Project (3GPP)}, {Technical Report (TR)} {RAN-80},
  Sep. 2020, {version 15.4.0}.

\bibitem{ali1998doppler}
I.~Ali, N.~Al-Dhahir, and J.~E. Hershey, ``Doppler characterization for
  {L}{E}{O} satellites,'' \emph{IEEE Transactions on Communications}, vol.~46,
  no.~3, pp. 309--313, 1998.

\bibitem{sharma2020system}
S.~K. Sharma, J.~Q. Borras, N.~Maturo, S.~Chatzinotas, and B.~Ottersten,
  ``System modeling and design aspects of next generation high throughput
  satellites,'' \emph{IEEE Communications Letters}, vol.~25, no.~8, pp.
  2443--2447, 2020.

\bibitem{khanzadi2014calculation}
M.~R. Khanzadi, D.~Kuylenstierna, A.~Panahi, T.~Eriksson, and H.~Zirath,
  ``Calculation of the performance of communication systems from measured
  oscillator phase noise,'' \emph{IEEE Transactions on Circuits and Systems I:
  Regular Papers}, vol.~61, no.~5, pp. 1553--1565, 2014.

\bibitem{javed2020journey}
S.~Javed, O.~Amin, B.~Shihada, and M.-S. Alouini, ``A journey from improper
  gaussian signaling to asymmetric signaling,'' \emph{IEEE Communications
  Surveys \& Tutorials}, vol.~22, no.~3, pp. 1539--1591, 2020.

\bibitem{9635690}
W.~Deng, Y.~Xia, Z.~Li, and W.~Pei, ``On the distribution of sinr for widely
  linear {M}{M}{S}{E} {M}{I}{M}{O} systems with rectilinear or
  quasi-rectilinear signals,'' \emph{IEEE Transactions on Vehicular
  Technology}, vol.~71, no.~2, pp. 1643--1655, 2022.

\bibitem{picinbono1995widely}
B.~Picinbono and P.~Chevalier, ``Widely linear estimation with complex data.''

\bibitem{jayaprakash2019analysis}
A.~Jayaprakash, H.~Chen, P.~Xiao, B.~G. Evans, Y.~Zhang, J.~Y. Li, and A.~B.
  Awoseyila, ``Analysis of candidate waveforms for integrated
  satellite-terrestrial 5{G} systems,'' in \emph{IEEE 2nd 5{G} World Forum
  (5GWF)}, 2019, pp. 636--641.

\bibitem{huang2021synchronization}
M.~Huang, J.~Chen, and S.~Feng, ``Synchronization for {O}{F}{D}{M}-based
  satellite communication system,'' \emph{IEEE Transactions on Vehicular
  Technology}, vol.~70, no.~6, pp. 5693--5702, 2021.

\bibitem{silva2005survey}
C.~P. Silva, C.~J. Clark, A.~A. Moulthrop, and M.~S. Muha, ``Survey of
  characterization techniques for nonlinear communication components and
  systems,'' in \emph{IEEE Aerospace Conference}, 2005, pp. 1713--1737.

\end{thebibliography}
%%%%%%%%%%%%%%%%%%%%%%%%%%%%%%%%%%%%%%%%%%%%%%%%%%%%%%%%%%%%%%%%%%%%%%%
%%%%%%%%%%%%%%%%%%%%%%%%%%%%%%%Bio%%%%%%%%%%%%%%%%%%%%%%%%%%%%%%%%%%%%%
%%%%%%%%%%%%%%%%%%%%%%%%%%%%%%%%%%%%%%%%%%%%%%%%%%%%%%%%%%%%%%%%%%%%%%%
\end{document}